\documentclass[12pt]{article} 
\hoffset = - 0.5 truecm \voffset=-1.4 truecm
\def\beq{\begin{equation}}   \def\eeq{\end{equation}}
\def\bea{\begin{eqnarray}}  \def\eea{\end{eqnarray}} 
\def\noi{\noindent} \def\beeq{\begin{eqnarray}}
\def\eeeq{\end{eqnarray}}
\def\lsim{\raise0.3ex\hbox{$<$\kern-0.75em\raise-1.1ex\hbox{$\sim$}}}
\def\gsim{\raise0.3ex\hbox{$>$\kern-0.75em\raise-1.1ex\hbox{$\sim$}}}

\usepackage{axodraw}
\usepackage{epsfig}

\begin{document}

\begin{titlepage}
 
\begin{flushright}
LAPTH-887/01\\
LPT-Orsay 01-105\\
IPPP/01/68\\
DCPT/01/136\\
December 2001
\end{flushright}
\vspace{1.cm}

\begin{center}
\vbox to 1 truecm {}
{\large \bf A NLO calculation of the large--p$_{\bf T}$} \par \vskip 3 truemm
{\large \bf  photon + photon $\to$ photon + jet cross section} 
\par \vskip 3 truemm
\vskip 1 truecm {\bf M. Fontannaz$^{(a)}$, J. Ph. Guillet$^{(b)}$, G. 
Heinrich$^{(c)}$} \vskip 3 truemm

{\it $^{(a)}$ Laboratoire de Physique Th\'eorique, UMR 8627 CNRS,\\ 
Universit\'e
Paris XI, B\^atiment 210, 91405 Orsay Cedex, France}\\ 

 \vskip 3 truemm

{\it $^{(b)}$ LAPTH, UMR 5108 du CNRS associ\'ee \`a l'Universit\'e 
de Savoie, \\ BP 110, Chemin de Bellevue, 74941 Annecy-le-Vieux 
Cedex, France} 

\vskip 3 truemm

{\it $^{(c)}$ Department of Physics, University of Durham, Durham DH1 3LE, England}

\vskip 2 truecm

\normalsize

\begin{abstract}

We study the production of an isolated large--$p_T$ photon as well as 
the production of an isolated prompt photon plus a jet in $e^+e^-$
collisions. Our results are obtained by a NLO Monte Carlo program of 
partonic event generator type. We discuss the possibilities to 
constrain the parton densities in the real photon  and compare 
to preliminary OPAL data.

\end{abstract}

\vspace{3cm}

\end{center}

\end{titlepage}

\baselineskip=22 pt
\section{Introduction}
\hspace*{\parindent} Hard reactions involving real photons, such as 
the photoproduction of
jets~\cite{1r,Frixione:1997ks,Klasen:1997it,Harris:1997hz,3r} 
or of large--$p_T$
photons~\cite{Gordon:1994sm,Gordon:1995km,Krawczyk,Fontannaz:2001ek,Fontannaz:2001nq}, 
are often presented as good tests of QCD. Indeed the real photon having a
pointlike coupling to a quark of the hard subprocess is a good probe of
the short distance dynamics. This would particularly be the case if one could
forget the hadronic component of real photons. In the initial state,
the incoming real photon can fluctuate into states made of quarks and
gluons which take part in the hard scattering process. This mechanism,
referred to as the resolved process, leads to the introduction of quark
and gluon distributions in real photons. Similarly a final state
large--$p_T$ photon can also be produced by the fragmentation of a
large--$p_T$ parton emerging from the hard scattering process\,; this
process is described by introducing the fragmentation function of
partons into real photons. Therefore the reactions involving real
photons are far from being simple processes and contain the whole
complexity of the pure hadronic collisions. This is particularly true for
the reaction $\gamma \gamma \to \gamma + X$ in which the final photon
is produced at large transverse momentum. Each of the photons can have
a direct or a resolved interaction such that eight different processes
contribute to the cross section. \par

However this complexity has a positive aspect if the various
contributions to the cross section can be disentangled by means, for 
instance, of kinematical cuts. In an ideal
situation it should be possible to separately observe direct
contributions which provide unambiguous tests of QCD, resolved
contributions leading to the measurement of the quark and gluon
distributions in the photon and fragmentation contributions giving
access to the quark and gluon fragmentation functions. These
fragmentation contributions can be strongly suppressed by using
isolation criteria \cite{5r,6r} which eliminate events with a too large
hadronic transverse energy in a cone surrounding the photon. For
instance the criterion used by the OPAL experiment \cite{7r}, that we
analyze in this paper, completely suppresses the fragmentation
contribution and the remaining contributions to the cross section are
due to direct and resolved incoming photons, the final photon being
always a direct one. \par

One way to disentangle the resolved
contribution from the direct one rests on the measurement of the
momentum of the parton entering the hard subprocess \cite{1r}
(a measurement which requires the observation of the whole final state). When
the ratio of this momentum to the incoming photon momentum is close to one,
one expects the direct contribution to be dominant (the photon momentum
is not shared among the quarks and gluons)\,; for a ratio smaller than one
the resolved contribution is the larger one. One interesting aspect of
the $\gamma + \gamma \to \gamma + X$ reaction at LEP II energies is
that the dominant contributions to the cross section contain at 
least one resolved photon in the initial state. The
direct-direct correction is negligible,
and the observation of the $\gamma + \gamma \to \gamma + X$ cross
section opens the possibility to measure the parton distributions in the
photon. \par

Another interesting point of the reaction $\gamma + \gamma \to \gamma +
X$ at LEP II energies is the small contribution of the NLO
corrections. As we shall see below, these corrections represent -- for the
scales set equal to $p_{T_{\gamma}}$ -- about 10 \% of the cross section.
This result shows that the QCD corrections are well under control.
Moreover the scale dependence is very weak and therefore the 
theoretical predictions are unambiguous. \par

The possibility to constrain the parton distributions in the photon, 
especially the rather poorly known gluon distribution, via the process 
$\gamma + \gamma \to \gamma +X$ has  
already been pointed out in~\cite{rohini,gs}. 
While \cite{rohini} is a leading order calculation, in \cite{gs} the 
calculation is done at next-to-leading order, but only the fully
inclusive cross section is calculated, without the possibility to 
consider isolated photons or a photon + jet final state. 

The plan of the paper is the following. Section 2 is devoted to the
theoretical framework and to the discussion of predictions for the
inclusive cross sections $d\sigma /dp_{T_{\gamma}}$ and
$d\sigma/d\eta_{\gamma}$. In Section 3 we study the reaction $\gamma +
\gamma \to \gamma + jet + X$, emphasizing the difference to the
inclusive case. Comparisons with preliminary results of the OPAL 
experiment are
performed in Section 4. Section 5 contains the conclusion.

\section{Theoretical Framework and Predictions}

\subsection{General Setting} \label{sec21}\hspace*{\parindent} 
As the general
framework of the calculation has already been described in detail in
\cite{8r,Fontannaz:2001ek,3r}, we will give only a brief overview on the method here. \par

In $e^+e^-$ reactions, the electrons can act like a source of quasi-real
photons whose spectrum can be described by the Weizs\"acker-Williams
formula

\beq \label{1e} f_{\gamma}^e(y) = {\alpha_{em} \over 2 \pi} \left \{ {1
+ (1 - y)^2 \over y} \ \ln \ {Q^2_{max} (1- y) \over m_e^2\ y^2} \ - \
{2(1 - y) \over y} \right \} \quad . \eeq

\noi This approximation is valid as long as $Q^2_{max}/p_T^2 \ll 1$
where $p_T$ is a large scale characterizing the hard subprocess, for
instance the transverse momentum of the final photon. In this section
we use this approximation, postponing to Section~4 a discussion of its
validity when comparing our predictions with OPAL data. \par

The quasi-real photon then either takes part {\it directly} in the hard
scattering process, or it acts as a composite object, being a source of
partons which take part in the hard subprocess. The latter mechanism is
referred to as {\it resolved} process and is parametrized by the photon
structure functions $F_{a/\gamma}(x_{\gamma}, Q^2)$. Thus the
distribution of partons in the electron is given by the convolution

\beq \label{2e} F_{a/e}(x_e, M) = \int_0^1 dy \ dx_{\gamma} \
f_{\gamma}^e (y) \ F_{a/\gamma} (x_{\gamma}, M) \delta (x_{\gamma} y -
x_e) \eeq

\noi where in the ``direct'' case $F_{a/\gamma} (x_{\gamma}, M) =
\delta_{a\gamma} \delta (1 - x_{\gamma})$. \par

Similarly, a high-$p_T$ photon in the final state can either originate
directly from the hard scattering process or it can be produced by the
fragmentation of a hard parton emerging from the hard scattering
process. However, as discussed in the introduction, this fragmentation
contribution can strongly be suppressed by using an isolation
criterion. It turns out that this isolation is also useful at the
experimental level in order to single out the prompt photon events from
the background of secondary photons produced by the decays of light
mesons. The isolation criterion used by the OPAL collaboration is the
one proposed by Frixione \cite{5r} which leads to a complete suppression
of the fragmentation contributions (contributions involving the
fragmentation function $D_a^{\gamma}(z, M^2)$ of parton $a$ into a
photon). At the parton level we have the following constraints~: for
each parton $i$, the distance

\beq \label{3e} R_{i\gamma} = \sqrt{(\phi_i - \phi_{\gamma})^2 +
(\eta_i - \eta_{\gamma})^2} \eeq

\noi to the photon is computed in $(\phi, \eta)$ space, where $\phi$ and
$\eta$ are the azimuthal angle and pseudorapidity, respectively. A
photon is kept if the condition

\beq \label{4e} \sum_i E_{T,i} \Theta (\delta - R_{i,\gamma}) \leq 0.2
\cdot E_{T, \gamma} {1 - \cos (\delta ) \over 1 - \cos (R)}\quad ,
\quad \hbox{for all $\delta \leq R$} \eeq

\noi is fulfilled, where $E_{T,i}$ is the transverse energy of the
$i^{th}$ parton, $\Theta$ is the step function, which ensures that only
particles in the cone with aperture $\delta$ contribute to
the sum, and the cone radius is $R = 1$. \par

As a result, we only have to consider contributions with a direct final
photon. These can be classified by the three production mechanisms illustrated in 
Fig.~1\,: 1) direct-direct, 2) single resolved, 3) double resolved. 
Note that by "single resolved" we denote the sum of direct-resolved 
and resolved-direct contributions. 

\begin{figure}[htb]
\begin{center}
\begin{picture}(400,150)(-40,-100)
\Photon(0,0)(-30,20){2}{5}
\ArrowLine(0,0)(30,20)
\Line(0,0)(0,-40)
\Photon(0,-20)(30,-20){2}{5}
\Photon(0,-40)(-30,-60){2}{5}
\ArrowLine(30,-60)(0,-40)
\Text(-5,-88)[t]{direct - direct}
\Photon(170,0)(140,20){2}{5}
\ArrowLine(170,0)(200,20)
\Line(170,0)(170,-40)
\ArrowLine(145,-55)(170,-40)
\Photon(170,-40)(200,-60){2}{5}
\BCirc(140,-57){6}
\Line(146,-58)(162,-50)
\Line(144,-53)(155,-43)
\Line(146,-60)(162,-54)
\Photon(105,-75)(135,-60){2}{4}
\Text(155,-88)[t]{direct - resolved}
\Photon(275,34)(306,20){2}{4}
\ArrowLine(340,0)(315,15)
\BCirc(310,17){6}
\Line(316,18)(332,10)
\Line(313,12)(325,3)
\Line(316,20)(332,14)
\Photon(340,0)(370,20){2}{5}
\Line(340,0)(340,-40)
\ArrowLine(315,-55)(340,-40)
\Gluon(340,-40)(370,-60){2}{5}
\BCirc(310,-57){6}
\Line(316,-56)(332,-50)
\Line(314,-53)(325,-43)
\Line(316,-59)(332,-54)
\Photon(275,-75)(305,-60){2}{4}
\Text(325,-88)[t]{resolved - resolved}
\end{picture}
\end{center}
\caption{Examples of contributing processes.}
\label{fig1}
\end{figure}
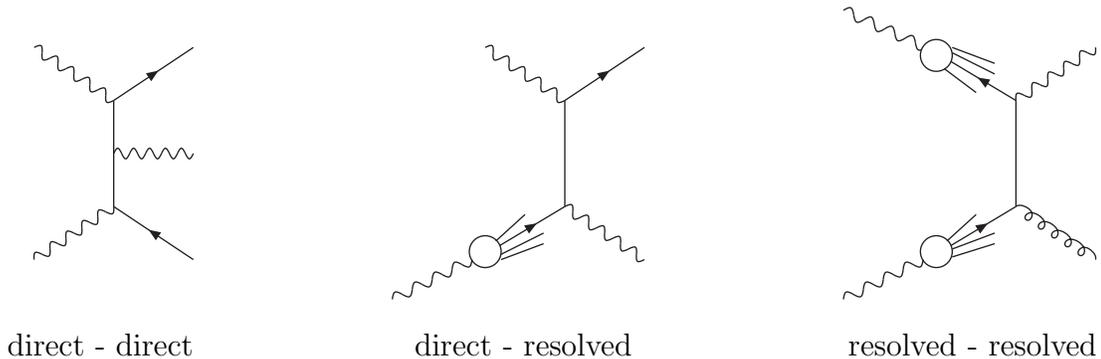


\noi We implemented the full set of next-to-leading order corrections to the
single resolved and double resolved processes. Note that the 
direct-direct process is by itself a higher order correction\,;
there is no Born contribution involving a $2 \to 2$ subprocess in this
reaction. The isolation criterion does not modify the leading order 
contributions for the direct-resolved and resolved-resolved case shown 
in Fig.~\ref{fig1}. But
at NLO a parton can be close to the photon in the $(\phi , \eta )$
plane and the cross section is modified by the constraint (\ref{4e}).
\par

In the calculation of the higher order corrections a phase-space slicing
method is used which allows to isolate and analytically calculate the divergent
infrared and collinear contributions. The non-divergent parts of the $2
\to 3$ contributions are calculated numerically. A partonic event
generator (allowing negative weights) has been built on these bases.
The interested reader may find a detailed description of this generator
in ref.~\cite{8r}. 

\subsection{Predictions}

Let us now study the NLO cross section of the $e^+e^- \to e^+e^-\gamma
X$ reaction obtained in the above theoretical framework and by using 
the OPAL kinematical conditions. Thus we use
$\sqrt{S_{e^+e^-}} = 196.6$~GeV, $Q^2_{max} = 10$~GeV$^2$ and $0 \leq y
\leq 1$. We also use the AFG parton distributions \cite{afg} and the value
$\Lambda_{\overline{MS}}^{(4)} = 300$~MeV. (We work with 4 flavours).
The factorization scale $M$ and the renormalization scale $\mu$ are set
equal to $p_{T_{\gamma}}$.
The NLO cross sections $d\sigma /dp_{T_{\gamma}}$ associated with the
different production mechanisms, integrated in the
rapidity range $- 1 \leq \eta_{\gamma} \leq 1$, are displayed in Fig.~2. 
 
\begin{figure}[htb]
\begin{center}
\mbox{\epsfig{file=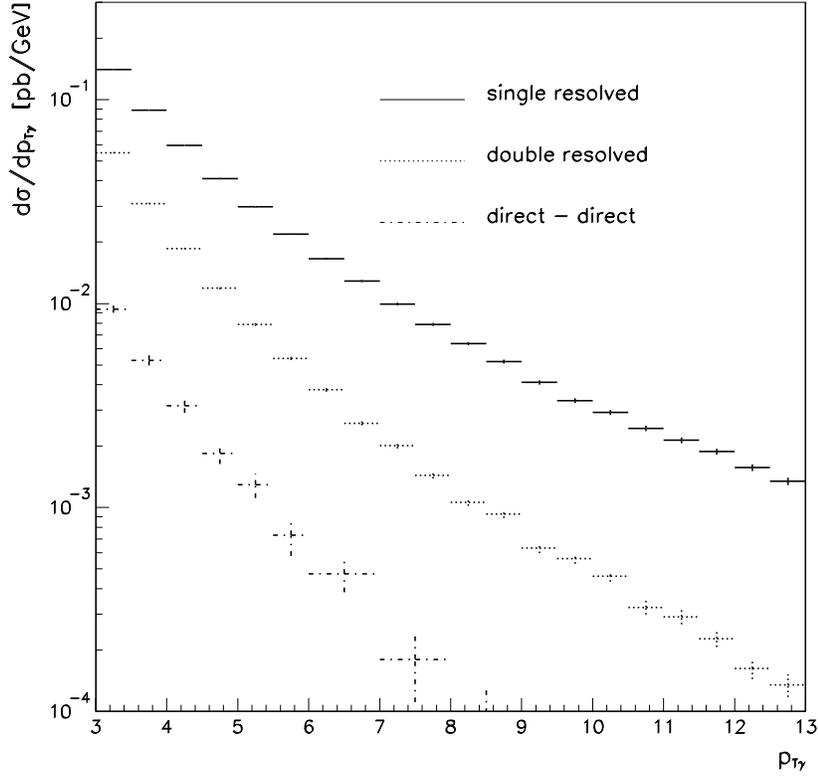,height=11.8cm}}
\end{center}
\caption{The NLO single resolved, 
double resolved and direct cross sections $d\sigma^{\gamma+X}/dp_{T_{\gamma}}$.} 
\label{fig2}
\end{figure}

For the
scales $\mu = M = p_{T_{\gamma}}$ used here the single resolved
contribution is larger than the double resolved contribution by a
factor 3 to 6 which increases when $p_{T_{\gamma}}$ increases. The single
resolved contribution constitutes about 70\,\% of the total cross section. 

\begin{figure}[htb]
\begin{center}
\mbox{\epsfig{file=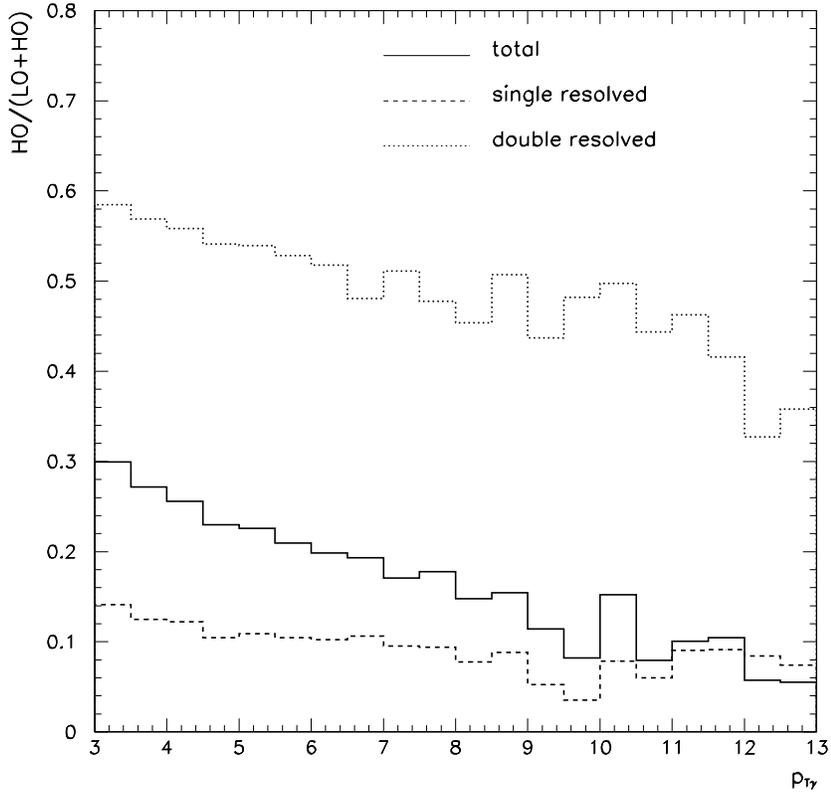,height=12cm}}
\end{center}
\caption{Ratios HO/(LO + HO) for
the $e^+e^- \to e^+e^-\gamma X$ cross section, calculated with the
scale $\mu = M = p_{T_{\gamma}}$, as a function of $p_{T_{\gamma}}$.}
\label{fig3}
\end{figure}

The ratio of the HO corrections to the NLO (= LO + HO) cross sections is
displayed in Fig.~\ref{fig3}.
The HO corrections are small for the single resolved cross section
(HO/(LO + HO) $\simeq 0.1$)) and large for the resolved-resolved case
(HO/(LO + HO) $\simeq 0.5$). The direct-direct HO contribution is
smaller than the single resolved HO contribution\,; it becomes negative at
large $p_{T_{\gamma}}$ ($p_{T_{\gamma}}\,\gsim \, 10$~GeV). On this occasion we
remind the reader that the individual contributions to the cross section
are not physical. They depend on the factorization and renormalization
scheme (here the $\overline{MS}$ scheme). For instance the
direct-direct HO contribution comes from the $\gamma + \gamma \to
\gamma + q + \bar{q}$ cross section from which the initial state collinear
divergences have been subtracted in the $\overline{MS}$-scheme. 
The finite remainders of the subtracted singularities 
depend strongly on the factorization scale $M$, this dependence 
being compensated only when adding the single resolved contributions.  

\begin{figure}[htb]
\begin{center}
\mbox{\epsfig{file=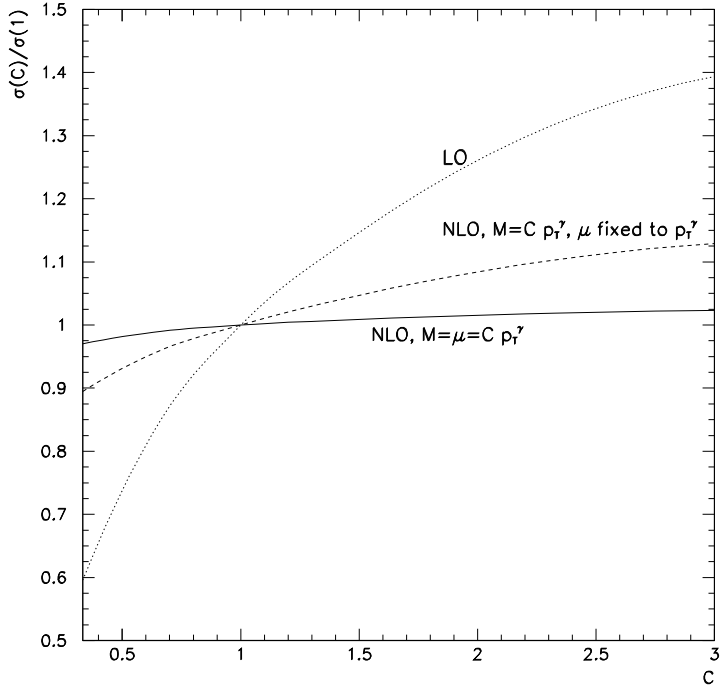,height=10.5cm}}
\end{center}
\caption{The dependence of $d\sigma^{\gamma +X} /dp_{T_{\gamma}}$ at 
$p_{T_{\gamma}}=6$ GeV on
the renormalization and factorization scales as a function of $C$. 
Solid line: $\mu = M = C \,p_{T_{\gamma}}$ (NLO), 
dotted line: $\mu = M = C \,p_{T_{\gamma}}$ (LO),
dashed line: $\mu$ fixed to $\,p_{T_{\gamma}}$, $M$ varied between 
$p_{T_{\gamma}}/3$ and $3\,p_{T_{\gamma}}$ (NLO).}
\label{fig4}
\end{figure}

Let us turn to the question of the stability of the predictions with
respect to the choice of the renormalization and factorization scales.
In Fig.~\ref{fig4} we display the variation of the cross section $d\sigma
/dp_{T_{\gamma}}$ calculated at $p_{T_{\gamma}} = 6$~GeV as a function
of the parameter $C$ which defines the scale choice, $\mu = M = C\,
p_{T_{\gamma}}$ resp. $M = C\,p_{T_{\gamma}}$\,; 
the cross sections are normalized to one at $C = 1$.
The increase of the leading order\footnote{What we denote by "leading order" 
only refers to the matrix elements. For the parton distributions we always 
use the NLL fits, and for $\alpha_s$ we use an exact solution of the 
two-loop renormalization group equation, and not an expansion in
$\log\mu/\Lambda$.} 
cross section with increasing scales 
reflects the fact that -- in contrast to the hadronic structure functions 
which show the well known scaling violations 
-- the photon structure functions grow uniformly 
like $\log (Q^2/\Lambda^2)$ for large $Q^2$ since the pointlike piece 
dominates at large $Q^2$.  

We note that the variations of the NLO cross section are smaller than 5\,\%,
when the values of the scales $M$ and $\mu$ are changed simultaneously 
by a factor 9. At the same
time, each individual cross section strongly varies. For instance the
direct-direct contribution varies from 7.85\,$\times 10^{-3}$\,pb 
to $-6.67\,\times 10^{-3}$\,pb
whereas the total cross section is equal to 
23.82\,$\times 10^{-3}$\,pb at $C = 1$.
If we vary the factorization scale $M$ only, keeping $\mu$ fixed to 
$p_{T_{\gamma}}$, the variations of the NLO cross section are 
of course larger
 since  cancellations of $\log{p_{T_{\gamma}}^2/M^2}$ 
 and $\log{\mu^2/p_{T_{\gamma}}^2}$ terms are spoiled,
 but they are still only of the order of 20\% for a 
variation of $M^2$ by almost two orders of magnitude, as can be seen from
Fig.~\ref{fig3}. 
\par


To conclude this short study of the cross section $d\sigma
/dp_{T_{\gamma}}$ we can say that the NLO calculation is well under
control\,; the HO corrections are small and the sensitivity of the cross
section to the scales is negligible. Therefore the reaction $e^+e^- \to
e^+e^-\gamma X$ offers interesting tests of QCD and should allow to put
constraints on the parton distributions in the real photon. \par

\begin{figure}[htb]
\begin{center}
\mbox{\epsfig{file=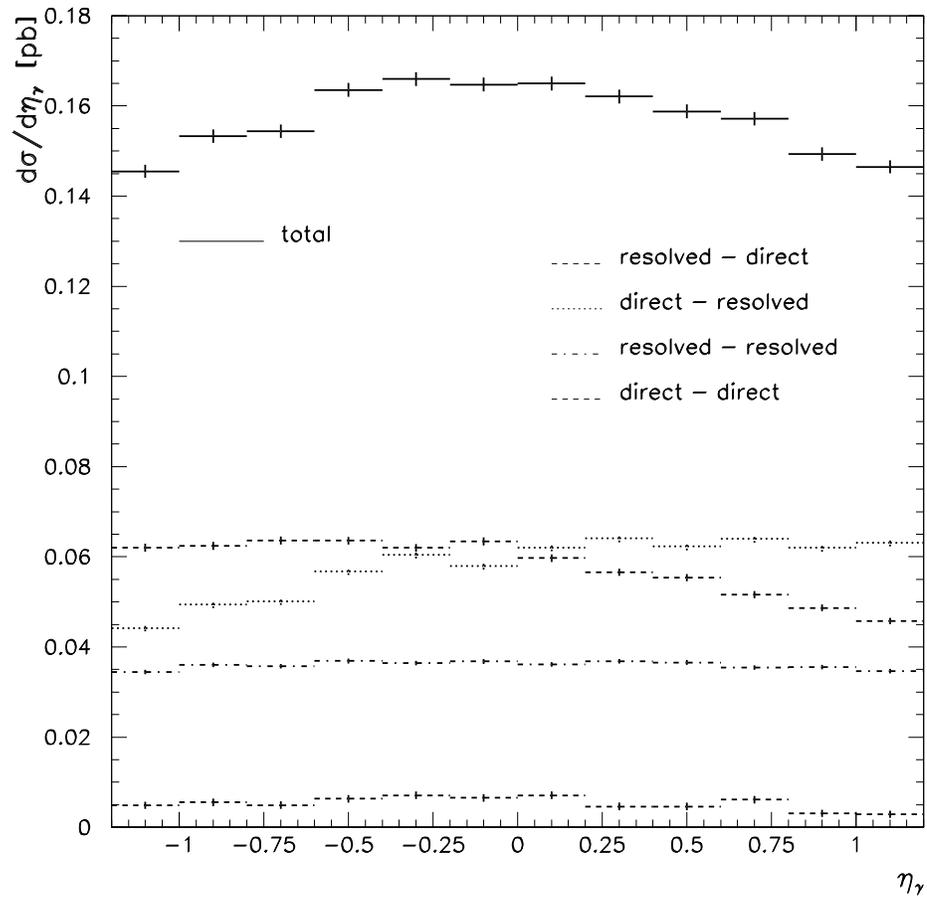,height=13.5cm}}
\end{center}
\caption{The cross section $d\sigma^{\gamma +X}/d\eta_{\gamma}$ as a 
function of $\eta_{\gamma}$ for $p_{T_{\gamma}} \geq 3$~GeV.}
\label{fig5}
\end{figure}

The cross section $d\sigma /d\eta_{\gamma}$ is a good observable to
determine $F_{a/\gamma}(x, M)$, because the value of $x$ is closely
related to the value of $\eta_{\gamma}$. As a consequence the shape of
$d\sigma /d\eta_{\gamma}$ gives indications on the $x$-dependence of
the parton distributions in the photon. The cross sections $d\sigma
/d\eta_{\gamma}$ are shown in Fig.~\ref{fig5} for 
$p_{T_{\gamma}} \geq 3$~GeV. Obviously the result is symmetric with respect to
$\eta_{\gamma}=0$ within the statistical errors.
The box contribution 
$\gamma g \to \gamma g$ to the single resolved part is of the order of 5\% of the 
total single resolved contribution, and of the same order of magnitude as 
the direct-direct contribution. However, in view of other uncertainties of the
order of 5\,\% (see Section 4), it has not been included in the results shown.
The box correction $g g \to \gamma g$ to the double resolved 
part is one order of magnitude smaller than the $\gamma g \to \gamma g$ box
and therefore completely negligible. 

Large values of $\eta_{\gamma}$ enhance the direct-resolved
contribution (where the incoming direct photon goes towards positive
rapidity). The resolved-resolved contribution is quite flat and the
direct-direct one almost negligible (at the scales $\mu = M =
p_{T_{\gamma}}$). 
A value of $p_{T_{\gamma}} ^{min}$ larger than 
3~GeV would enhance the single resolved cross section with respect 
to the double resolved one.

\clearpage

\section{Photon--jet cross section} \hspace*{\parindent}
In the introduction, we discussed the possibility to enhance a resolved contribution
with respect to a direct one by fixing the momentum of the partons
entering the hard subprocess. Following the OPAL collaboration, we
define

\beq \label{5e} x^{\pm} = {(E_{jet} \pm p_z^{jet} ) + (E_{\gamma} \pm
p_z^{\gamma}) \over \sum\limits_{had, \gamma} (E \pm p_z)} = {p_T^{jet}
\ e^{\pm \eta_{jet}} + p_T^{\gamma} \ e^{\pm \eta_{\gamma}} \over
y^{\pm} \ \sqrt{S_{e^+e^-}}} \eeq

\noi where $y^{\pm}$ are the fractional momenta of the quasi-real
initial photons oriented towards the positive and negative $z$-axis,
$y=E_{\gamma}/E_e$. The conventions are such that particle one
travels towards the positive $z$-direction, 
such that 
e.g. direct-resolved means that the direct photon -- with momentum 
fraction $x^+_{\gamma}$ -- moves to positive $z$ whereas the resolved photon 
(with momentum fraction $x^-_{\gamma}$) moves to negative $z$.  
The variables $x^{\pm}$ defined by eq.~(\ref{5e}) are exactly the longitudinal
momentum fractions (with respect to the photon momenta) of the partons entering
the hard subprocess if the latter is a $2 \to 2$ process. 
In the case of a $2 \to 3$ subprocess, these momentum fractions 
are no more fixed by the observation of the photon and a jet since there
are configurations with 
a third (unobserved) particle in the final state such that the 
true value of $x_{\gamma}^{\pm}$ in $F_{a/\gamma} (x_{\gamma}^{\pm}, M)$
(see eq.~(\ref{2e})) can be larger than $x^{\pm}$.   
Nevertheless $x^{\pm}$ remain useful
variables to constrain the parton momenta. 

The definition
(\ref{5e}) implies to measure the jet rapidity and transverse energy.
In ref.~\cite{3r} we defined another variable which does not require
the measurement of $p_T^{jet}$,
\beq \label{6e} x_{LL}^{\pm} = {p_T^{\gamma} (e^{\pm \eta_{jet}} +
e^{\pm \eta_{\gamma}}) \over y^{\pm} \ \sqrt{S_{e^-e^+}}} \quad . \eeq

This is certainly an advantage with respect to expression (\ref{5e}),
because the measurement of a jet transverse energy at small $p_T^{jet}$
($p_T^{jet} \sim 5$~GeV) can be very difficult and inaccurate\,; it is
delicate to disentangle the jet energy from the underlying event
energy. Moreover the variables (\ref{6e}) lead to smoother theoretical
distributions in the regions $x_{(LL)}^{\pm} \simeq 1$, which may be
useful if one wants to compare theory with data for these values of
$x_{(LL)}^{\pm}$. These points have been discussed in more detail in ref.
\cite{Fontannaz:2001nq} to which we refer the interested reader. 

\begin{figure}[htb]
\begin{center}
\mbox{\epsfig{file=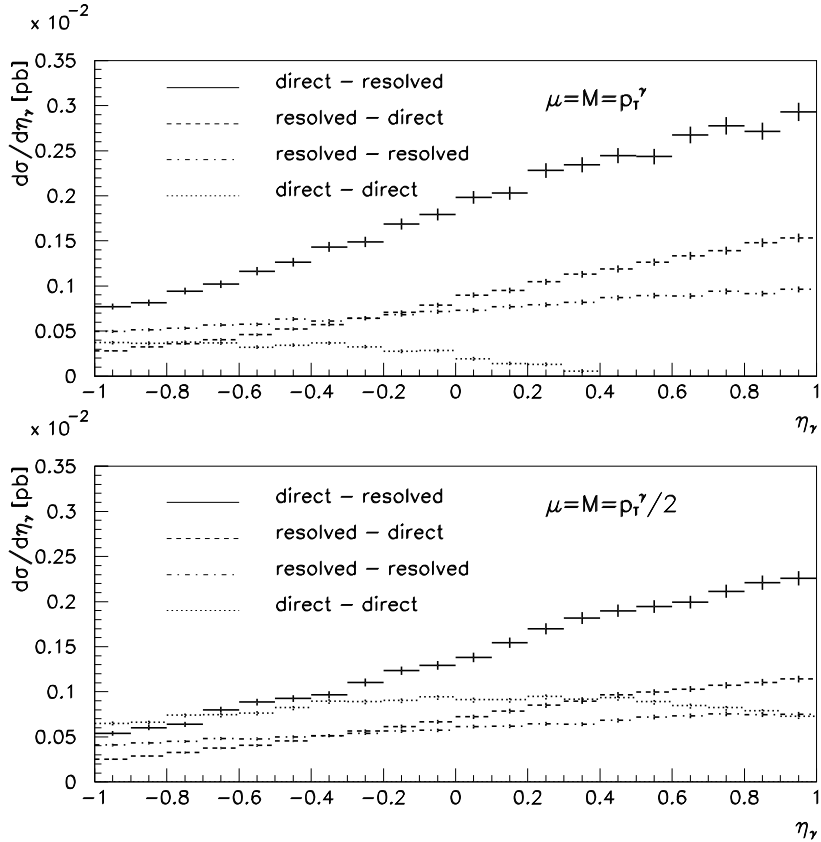,height=12cm}}
\end{center}
\caption{The cross sections
$d\sigma^{\gamma+jet}/d\eta_{\gamma}$ at $p_{T_{\gamma}} = 5$\,GeV and $1 \leq
\eta_{jet} \leq 2$ for two different scale choices.}
\label{fig6}
\end{figure}

Before using these variables, let us consider the behaviour of the
photon-jet cross section with respect to the photon and jet
rapidities. In order to enhance the direct-resolved contribution, 
we choose a large value of $\eta_{jet}$, $1 \leq \eta_{jet} \leq 2$, and
we vary $\eta_{\gamma}$. The jets are defined in agreement with the
cone algorithm used by the OPAL collaboration \cite{7r}, with a cone
radius $R = 1$ and a cut on the minimum value of $E_T^{jet}$ that we
fix equal to 3~GeV~\cite{opal1}. We also fix $p_T^{\gamma} = 5$~GeV. 
Our results are shown in Fig.~\ref{fig6} and they can be compared to those
displayed in Fig.~\ref{fig5}. 
\clearpage

We recall that, as always, the individual contributions to the total 
cross section have no physical meaning. Their relative importances are scale
dependent. 
However, choosing e.g. the scales $\mu=M=p_{T_{\gamma}}/2$ 
does not modify the fact that the direct-resolved contribution
dominates in the range $\eta_{\gamma}\,\gsim -0.7 $.
Fig.~\ref{fig6} shows the enhanced direct-resolved
contribution, corresponding to small values of $x_{(LL)}^-$ and large
values of $x_{(LL)}^+$ in expressions (\ref{5e}) or (\ref{6e}), for the 
two scale choices $\mu=M=p_{T_{\gamma}}$ and $\mu=M=p_{T_{\gamma}}/2$. 
Hence, by working at large photon and jet rapidities, we can enhance
the  direct-resolved contribution and study its behaviour as a function of 
$x_{(LL)}^-$. 
\begin{figure}[htb]
\begin{center}
\mbox{\epsfig{file=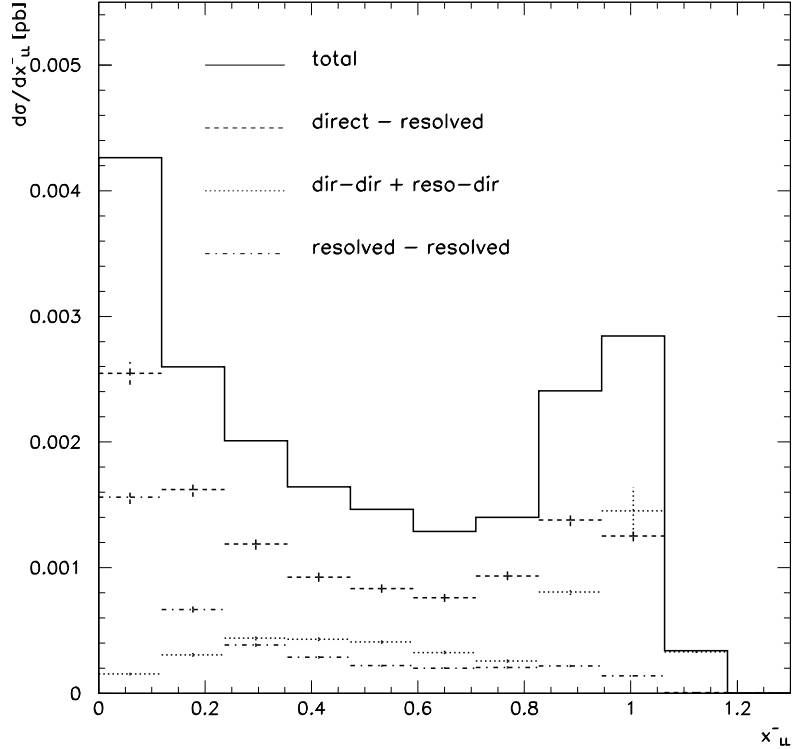,height=11.5cm}}
\end{center}
\caption{$d\sigma^{\gamma+jet}/dx_{LL}^-$  with
$p_{T_{\gamma}} = 5$\,GeV, $1 \leq \eta_{jet} \leq 2$ and $0.5 \leq
\eta_{\gamma} \leq 1$.}
\label{fig7}
\end{figure}

\begin{figure}[htb]
\begin{center}
\mbox{\epsfig{file=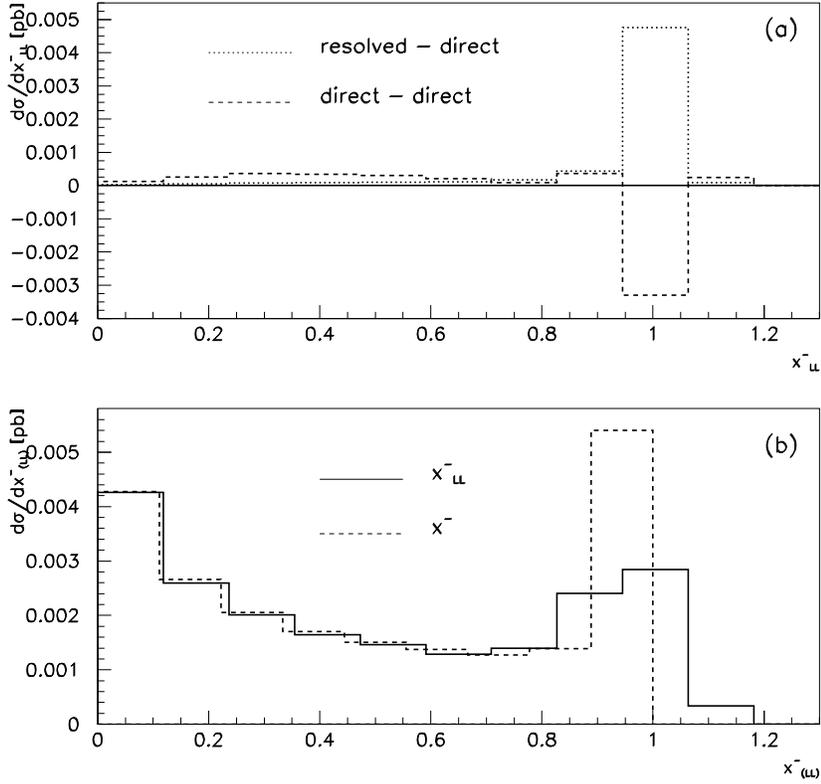,height=12cm}}
\end{center}
\caption{$d\sigma^{\gamma+jet}/dx_{(LL)}^-$  with
$p_{T_{\gamma}} = 5$\,GeV, $1 \leq \eta_{jet} \leq 2, 0.5 \leq
\eta_{\gamma} \leq 1$. \qquad\qquad\qquad
(a) cancellations between resolved-direct and direct-direct subprocesses
\hspace*{1.3cm}\,\quad
(b) comparison between $d\sigma^{\gamma+jet}/dx_{LL}^-$ and 
$d\sigma^{\gamma+jet}/dx^-$ (for the sum of all subprocesses). 
The distribution in $x_{LL}$ shows a much smoother behaviour at $x\simeq 1$.}
\label{fig7a}
\end{figure}

In Fig.~\ref{fig7} we display the distribution $d\sigma/dx_{LL}^-$ obtained with
$p_{T_{\gamma}} = 5$\,GeV, $1 \leq \eta_{jet} \leq 2$ and $0.5 \leq
\eta_{\gamma} \leq 1$. The contributions from direct-direct and 
resolved-direct have been summed up in Fig.~\ref{fig7} since at $x^-\simeq 1$ 
there are large cancellations between these two contributions due to the finite
remainders of the initial state collinear singularities, as has been explained
in Section~\ref{sec21}. This is illustrated in Fig.~\ref{fig7a}(a). 
Of course there are also cancellations between the direct-resolved and the
direct-direct part, but those are of major importance in the $x^+$ spectrum for 
$x^+\simeq 1$, affecting the  $x^-$ distribution only marginally. 

Thus we can conclude from Fig.~\ref{fig7}  that in the domain 
$x_{LL}^- \,\lsim \,0.8$ 
the cross section $d\sigma^{\gamma+jet}/dx_{LL}^-$ is quite sensitive to the parton
distributions 
$F_{a/\gamma}(x^-_{\gamma}\sim x_{LL}^-,p_{T_{\gamma}}) $ in the photon.
 
Until now we have used the variable $x_{LL}$ to study the various contributions
to the cross section $d\sigma/dx$. The use of $x^{\pm}$ as defined in
eq.~(\ref{5e}) leads to a similar result, except around $x\simeq 1$ where 
the cross section $d\sigma/dx_{LL}$ has a smoother behaviour, as illustrated 
in Fig.~\ref{fig7a}\,(b).

\begin{figure}[htb]
\begin{center}
\mbox{\epsfig{file=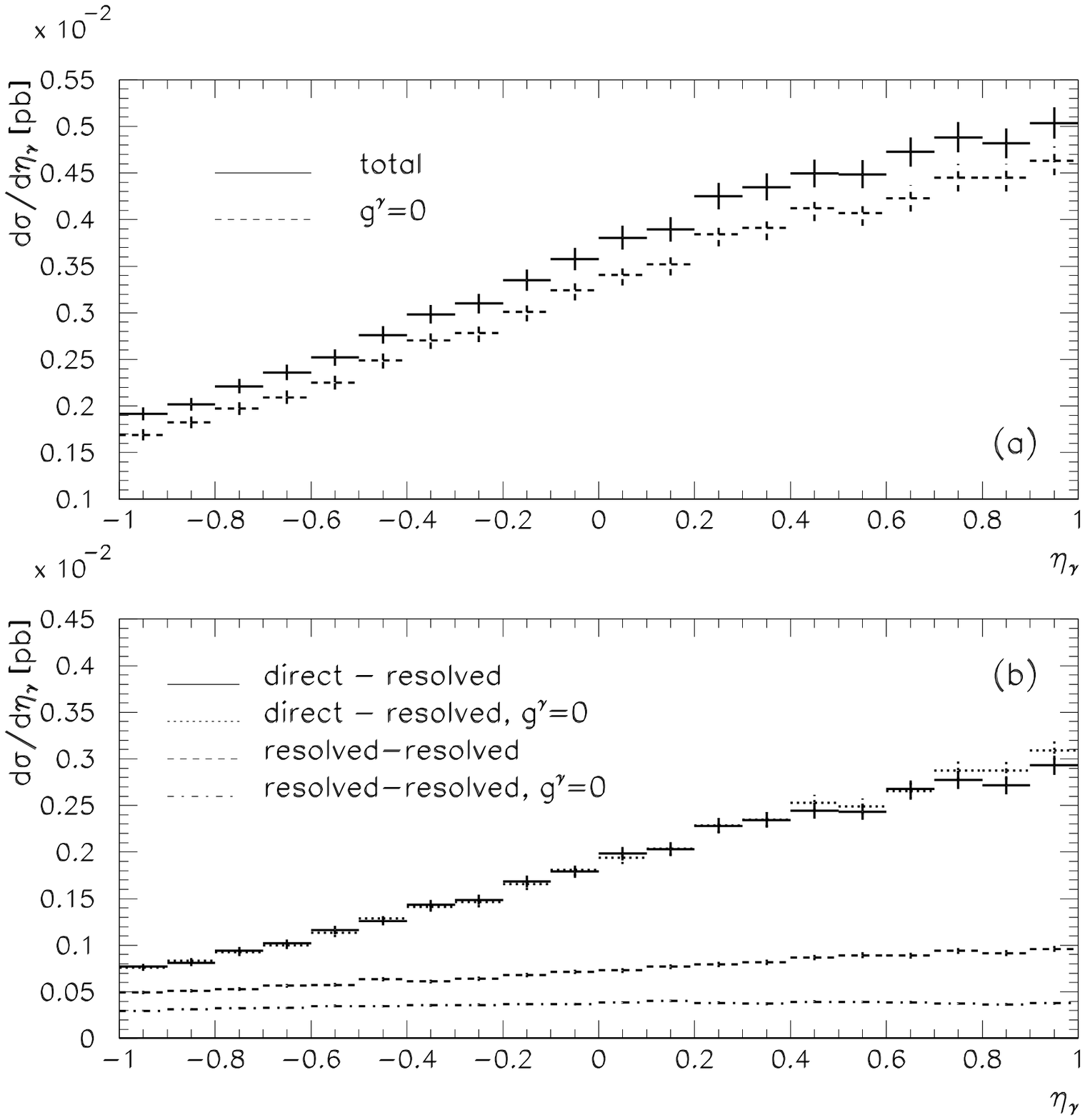,height=13cm}}
\end{center}
\caption{The contribution of the gluon in the resolved photon to the cross section
$d\sigma^{\gamma+jet}/d\eta_{\gamma}$ at $p_{T_{\gamma}} = 5$~GeV and $1 \leq
\eta_{jet} \leq 2$.}
\label{fig6b}
\end{figure}

Fig.~\ref{fig6b} shows that the contribution from the subprocess 
$\gamma + g \to q + \gamma + \bar{q}$ contributing to the direct-resolved part 
of $d\sigma/d\eta_{\gamma}$ is very small, and even negative for 
$\eta_{\gamma} > 0$,
such that the full direct-resolved contribution is slightly {\it lower} 
than the one where $g^{\gamma}$ has been set to zero. Thus the sensitivity 
to the gluon distribution in the photon as shown in Fig.~\ref{fig6b}(a)
is basically due to the double resolved part 
in the present study. 

\clearpage

\section{Comparison with OPAL data} \hspace*{\parindent} 
Before comparing theoretical predictions and data we have to assess the
validity of the Weizs\"acker-Williams (WW) approximation embedded in
formula (\ref{1e}). This approximation is obviously correct as long as
the photon virtuality is negligible with respect to the scale involved
in the hard subprocess, namely $p_{T_{\gamma}}$. But in OPAL $Q^2_{max}
\simeq 10$~GeV$^2$ is of the order of the subprocess scale,
$p_{T_{\gamma}} \geq 3$~GeV, and higher twist contributions
proportional to $Q_{max}^2/p_{T_{\gamma}}^2$ are not a priori
negligible. \par

However we have to note that the Weizs\"acker-Williams formula
is dominated by the term $\sim \log({Q^2_{max}/m_e^2})$ which is very 
large ($\sim 27$)
because of the small mass of the electron\,; therefore the WW
approximation should not be too bad, even if
$Q_{max}^2/p_{T_{\gamma}}^2 \simeq 1$. \par

Moreover the large virtuality of the incoming photon questions the use
of parton distributions in real photons. To estimate the importance of
the corrections due to the photon virtuality, let us consider the
perturbative quark distribution in a real photon in the Born
approximation

\beq \label{7e} q_{\gamma}(x,M) = 3\,{\alpha \over 2 \pi} (x^2 + (1 -
x)^2) \int_{Q_0^2}^{M^2} {dk^2 \over k^2} \quad . \eeq

The lower limit $Q_0^2$ is the boundary below which a perturbative
approach is no more valid. The value $Q_0^2 \simeq m_{\rho}^2 \simeq 
0.5$~GeV$^2$ leads to a photon structure function $F_2^{\gamma}(x, 
Q^2) =
 x\, \sum\limits_q e_q^2\, [q_{\gamma} (x, Q) + \bar{q}_{\gamma} (x, Q)]$ in
agreement with DIS $\gamma^{\star}\gamma$ experiments \cite{afg}. 
For a virtual
photon, $Q_0^2$ must be replaced by $max(Q^2 , Q_0^2)$ and the
convolution with the spectrum of the photon emitted by the electron
takes the form

\beq \label{8e} I = \int_{m_e^2}^{Q_{max}^2} {dQ^2 \over Q^2}
\int_{max(Q^2, Q_0^2)}^{M^2} {dk^2 \over k^2} \eeq

\noi in which we drop all $x$-dependence. If $Q^2>Q_0^2$, 
expression (\ref{8e}) can be
written as the sum of the real photon expression and a correction

\beq \label{9e} I = \int_{m_e^2}^{Q_{max}^2} {dQ^2 \over Q^2}
\int_{Q_0^2}^{M^2} {dk^2 \over k^2} - {1 \over 2} \ {\rm Log}^2 \ 
{Q_{max}^2 \over Q_0^2} \quad . \eeq

\noi For $Q_{max}^2 = 10$~GeV$^2$, $Q_0^2 = 0.5$~GeV$^2$ and $M^2 =
p_{T_{\gamma}}^2 = 25$~GeV$^2$, the correction is of the order of 5\,\%.
A similar estimation can be done for the non-perturbative part of the
photon structure function with a correction of the order of 15\,\%
which is also negative~\cite{japon}.

The conclusion of this study of the effects of the initial photon 
virtualities is that the WW approximation is valid in the OPAL
kinematical range up to corrections of a few tens of percents.

Now we compare our results to OPAL preliminary data~\cite{7r}. 
It has to be noted that in the experimental analysis the aim was
to keep only events from single and double resolved processes. 
Therefore a component identified as stemming mainly from direct-direct events 
(denoted by "final state radiation" (FSR) in~\cite{7r})  has been subtracted 
from the data. 
In a NLO calculation the final state radiation associated with the subprocess 
$\gamma\gamma\to\gamma q \bar{q}$ 
is a genuine contribution to the higher order corrections and thus cannot
be removed from the total cross section. It can only be suppressed by an
isolation criterion as discussed in Section~\ref{sec21}. 
In the present study the isolated direct-direct contribution is small, 
such that it should not disturb the comparison with the preliminary 
OPAL data which have large experimental errors. 

\begin{figure}[htb]
\begin{center}
\mbox{\epsfig{file=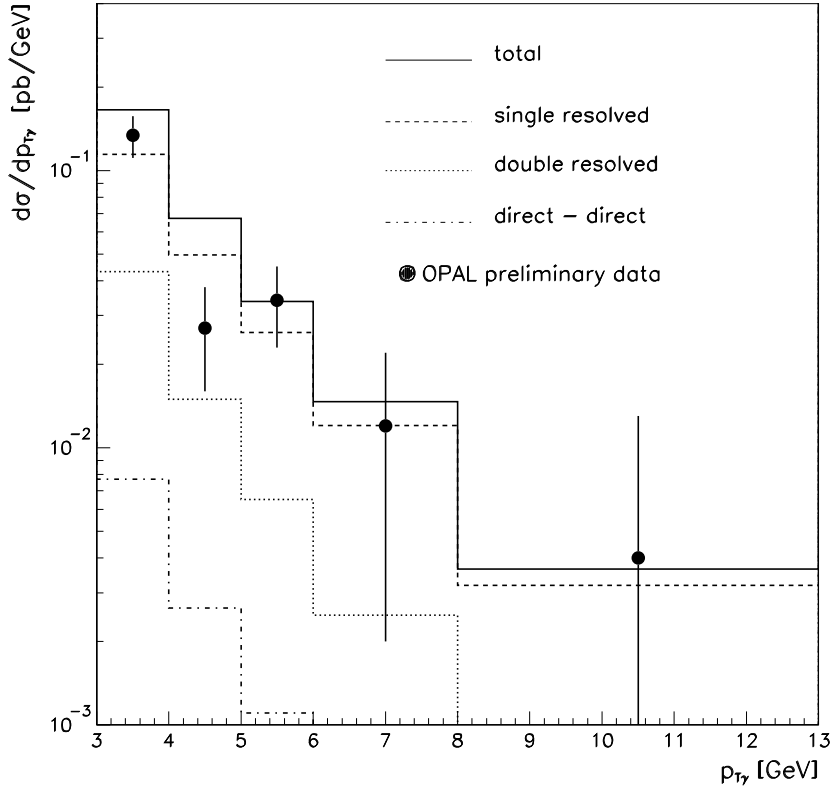,height=12cm}}
\end{center}
\caption{Comparison to preliminary OPAL data for 
$d\sigma^{\gamma+X}/dp_{T_{\gamma}}$ integrated over photon 
rapidities in the range $-1<\eta_{\gamma}<1$, at the scales 
$\mu=M=p_{T_{\gamma}}$.} 
\label{fig8}
\end{figure}
 
\begin{figure}[htb]
\begin{center}
\mbox{\epsfig{file=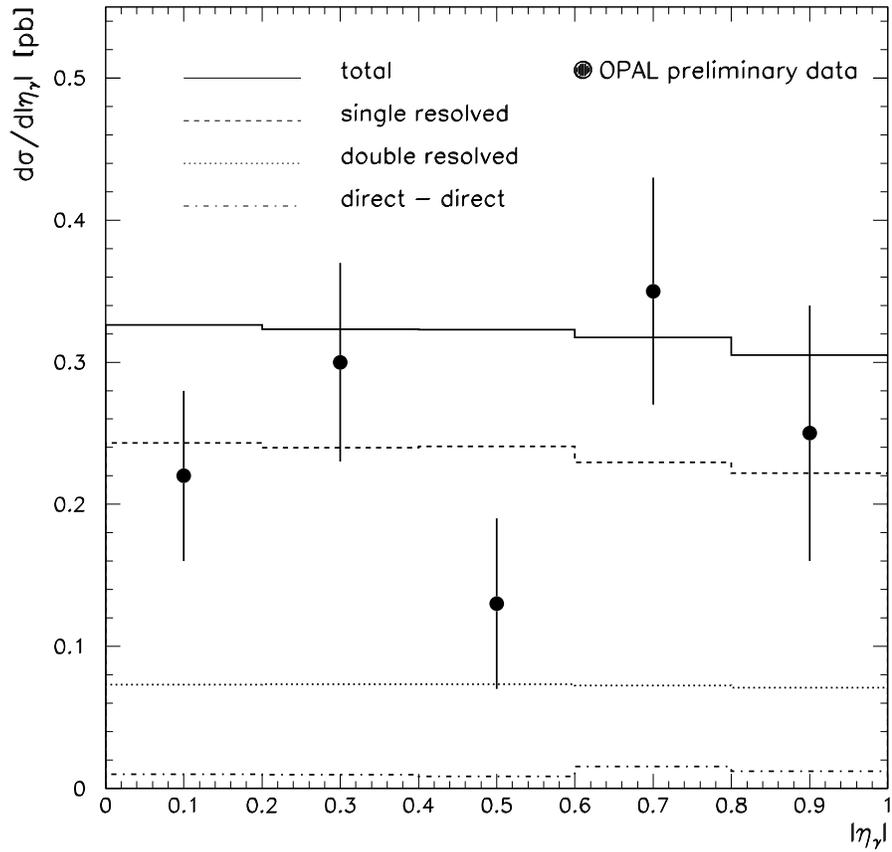,height=13cm}}
\end{center}
\caption{Comparison to preliminary OPAL data for
$d\sigma^{\gamma+X}/d|\eta_{\gamma}|$, at the scales 
$\mu=M=p_{T_{\gamma}}$.} 
\label{fig9}
\end{figure}

In Fig.~\ref{fig8} the cross section $d\sigma^{\gamma+X}/dp_{T_{\gamma}}$ is shown. 
One observes that NLO theory is somewhat higher than the data,  
in particular at low $p_{T_{\gamma}}$, but a correction term due to the  
rather large value of $Q^2_{max}$ would decrease the theory prediction (see
eq.~(\ref{9e})), and including the FSR in the experimental 
analysis would increase the data. On the other hand, it is hard to 
estimate if higher twist effects (which should be non-negligible only at low 
$p_{T_{\gamma}}$) would increase or decrease the theory prediction in the first 
$p_{T{\gamma}}$-bin.  \\
In short, taking into account a theoretical uncertainty of about 10\%, 
the agreement with the data is completely satisfactory. 
The comparison to  the rapidity distribution 
$d\sigma^{\gamma+X}/d|\eta_{\gamma}|$ is displayed in Fig.~\ref{fig9}.

In~\cite{7r} it is 
shown that PYTHIA reproduces the shapes of the distributions well, but has to be
scaled up by a factor 1.85 to be consistent with the data. 
One reason for this difference in normalisation 
might be the use of different parton distribution functions for the photon 
(to obtain the PYTHIA result, SAS-1D\,\cite{sas} -- which are LO fits --
have been used). \\
We further note that the data are also compatible with our "leading order" 
result. This does not come as a surprise since the 
higher order corrections are small and the preliminary data fluctuate a lot, 
but we recall that by "leading order" we only refer to the
matrix elements and not to the parton distribution functions and 
 $\alpha_s$.

\clearpage

\section{Conclusions} 
We have presented a complete NLO program for the calculation of the 
reactions $e^+e^-\to e^+e^-\,\gamma \,X$ and 
$e^+e^-\to e^+e^-\,\gamma + jet+X$, where the prompt photon is isolated
using the criterion of S.~Frixione~\cite{5r}. 
The higher order corrections turned out to be 
of the order of 20\% and the scale dependence is very weak: The total 
cross section varies by less than 5\% if the scales are 
varied between $\mu=M=p_{T_{\gamma}}/3$ and $3\,p_{T_{\gamma}}$. 

We discussed the possibility to constrain the parton distributions in 
the photon. For the photon--jet cross section, 
we have shown that 
one can enhance the direct-resolved component of the cross section by 
restricting the photon and jet rapidities to large values. 

Further, we discussed the validity of the Weizs\"acker-Williams approximation 
for a comparison to OPAL data with a rather large maximal photon virtuality 
of $Q^2_{\rm max}=10$\,GeV$^2$. The theoretical uncertainties 
introduced by using the parton distributions of real photons are also 
estimated in this context.
  
We compared the photon $p_T$ and rapidity distribution for the 
$\gamma\gamma\to \gamma+X$ cross section 
to preliminary OPAL data~\cite{7r}. In view of the large experimental errors, 
the agreement is good. 

\vspace*{8mm}       

\noindent{\bf \large Acknowledgements}


\noindent We would like to thank Stefan S\"oldner-Rembold from the OPAL collaboration
for providing us the preliminary data and for helpful comments. 
This work was supported by the EU Fourth Training Programme  
''Training and Mobility of Researchers'', network ''Quantum Chromodynamics
and the Deep Structure of Elementary Particles'',
contract FMRX--CT98--0194 (DG 12 - MIHT).



\begin{thebibliography}{99}

\bibitem{1r} ZEUS collaboration, J. Breitweg et al., Eur. Phys. J 
{\bf C11} (1999) 35;\\
H1 collaboration, C. Adloff et al., Eur. Phys. J {\bf 
C1} (1998) 97.

\bibitem{Frixione:1997ks}
S.~Frixione and G.~Ridolfi,
Nucl.\ Phys.\ B {\bf 507} (1997) 315.

\bibitem{Klasen:1997it}
M.~Klasen and G.~Kramer,
Z.\ Phys.\ C {\bf 76} (1997) 67;\\
%
M.~Klasen, T.~Kleinwort and G.~Kramer,
Eur.\ Phys.\ J.\ direct C {\bf 1} (1998) 1.

\bibitem{Harris:1997hz}
B.~W.~Harris and J.~F.~Owens,
Phys.\ Rev.\ D {\bf 56} (1997) 4007.

\bibitem{3r}P.~Aurenche, L.~Bourhis, M.~Fontannaz and J.~Ph.~Guillet,
Eur.\ Phys.\ J.\ C {\bf 17} (2000) 413.

\bibitem{Gordon:1994sm}
L.~E.~Gordon and J.~K.~Storrow,
Z.\ Phys.\ C {\bf 63} (1994) 581.

\bibitem{Gordon:1995km}
L.~E.~Gordon and W.~Vogelsang,
Phys.\ Rev.\ D {\bf 52} (1995) 58;\\
%
L.~E.~Gordon,
Phys.\ Rev.\ D {\bf 57} (1998) 235.

\bibitem{Krawczyk}
M.~Krawczyk, A.~Zembrzuski,
hep-ph/9810253 and hep-ph/0105166.

\bibitem{Fontannaz:2001ek}
M.~Fontannaz, J.~Ph.~Guillet and G.~Heinrich,
Eur.\ Phys.\ J.\ C {\bf 21} (2001) 303.

\bibitem{Fontannaz:2001nq}
M.~Fontannaz, J.~Ph.~Guillet and G.~Heinrich,
hep-ph/0107262, accepted for publication in Eur.\ Phys.\ J.\ {\bf C}.

\bibitem{5r} S.~Frixione, Phys. Lett. {\bf B429} (1998) 369.

\bibitem{6r} S. Catani, M. Fontannaz, J. Ph. Guillet and E. Pilon, in 
preparation.

\bibitem{7r} OPAL collaboration, OPAL Physics Note PN492.

\bibitem{rohini} M.~Drees and R.~M.~Godbole,
Phys.\ Lett.\ B {\bf 257} (1991) 425\,;\\
J.\ Phys.\ G {\bf 21} (1995) 1559.

\bibitem{gs} L.~E.~Gordon and J.~K.~Storrow,
Phys.\ Lett.\ B {\bf 385} (1996) 385.



\bibitem{8r} T. Binoth, J. Ph. Guillet, E. Pilon, M. Werlen, Eur. 
Phys. J {\bf C16} (2000) 311.

\bibitem{afg} P. Aurenche, J. Ph. Guillet and M. Fontannaz, Z. Phys. 
{\bf C64} (1994) 621.

\bibitem{opal1} 
G.~Abbiendi {\it et al.}  [OPAL Collaboration],
Eur.\ Phys.\ J.\ C {\bf 10} (1999) 547.

\bibitem{japon}
P.~Aurenche, J.~Ph.~Guillet, M.~Fontannaz, Y.~Shimizu, J.~Fujimoto and K.~Kato,
Prog.\ Theor.\ Phys.\  {\bf 92} (1994) 175.

\bibitem{sas}
G.~A.~Schuler and T.~Sjostrand,
Z.\ Phys.\ C {\bf 68} (1995) 607.

\end{thebibliography}
\end{document}